\documentclass[sigconf]{acmart}
\AtBeginDocument{%
  }

\copyrightyear{2025}
\acmYear{2025}
\setcopyright{cc}
\setcctype{by}
\acmConference[UIST '25]{The 38th Annual ACM Symposium on User Interface Software and Technology}{September 28-October 1, 2025}{Busan, Republic of Korea}
\acmBooktitle{The 38th Annual ACM Symposium on User Interface Software and Technology (UIST '25), September 28-October 1, 2025, Busan, Republic of Korea}\acmDOI{10.1145/3746059.3747615}
\acmISBN{979-8-4007-2037-6/2025/09}

\usepackage{amsbsy}
\usepackage{siunitx}
\usepackage{colortbl,array,xcolor}
\definecolor{myblue}{rgb}{0.5647,0.6784,0.8824}

\newcommand{\myTableCell}[1]{%
  {\renewcommand{\arraystretch}{1.0}%
  \begin{tabular}[c]{@{}l@{}}#1\end{tabular}}%
}

\begin{document}

\title[picoRing \textit{mouse}]{picoRing \textit{mouse}: ultra-low-power ring-based wireless tinymouse}


\author{Yifan Li}
\affiliation{%
  \institution{The University of Tokyo}
  \city{Tokyo}
  \country{Japan}
  }
\email{yifan217@akg.t.u-tokyo.ac.jp}
\orcid{0009-0005-9261-6391}

\author{Masaaki Fukumoto}
\affiliation{%
  \institution{Microsoft Corporation}
  \city{Beijing}
  \country{China}
  }
\email{fukumoto@microsoft.com}
\orcid{0009-0004-9368-7434}

\author{Mohamed Kari}
\affiliation{%
  \institution{Princeton University}
  \city{Princeton}
  \country{USA}
  }
\email{mokari@princeton.edu}
\orcid{0000-0003-4664-9983}

\author{Shigemi Ishida}
\affiliation{%
  \institution{Future University Hakodate}
  \city{Hokkaidō}
  \country{Japan}
  }
\email{ish@fun.ac.jp}
\orcid{0000-0003-0166-3984}

\author{Akihito Noda}
\affiliation{%
  \institution{Kochi University of Technology}
  \city{Kochi}
  \country{Japan}
  }
\email{noda.akihito@kochi-tech.ac.jp}
\orcid{0000-0002-6393-3196}

\author{Tomoyuki Yokota}
\affiliation{%
  \institution{The University of Tokyo}
  \city{Tokyo}
  \country{Japan}
  }
\email{yokota@ntech.t.u-tokyo.ac.jp}
\orcid{0000-0003-1546-8864}

\author{Takao Someya}
\affiliation{%
  \institution{The University of Tokyo}
  \city{Tokyo}
  \country{Japan}
  }
\email{someya@ee.t.u-tokyo.ac.jp}
\orcid{0000-0003-3051-1138}

\author{Yoshihiro Kawahara}
\affiliation{%
  \institution{The University of Tokyo}
  \city{Tokyo}
  \country{Japan}
  }
\email{kawahara@akg.t.u-tokyo.ac.jp}
\orcid{0000-0002-0310-2577}

\author{Ryo Takahashi}
\authornote{Correspondence to Ryo Takahashi}
\affiliation{%
  \institution{The University of Tokyo}
  \city{Tokyo}
  \country{Japan}
  }
\email{takahashi@akg.t.u-tokyo.ac.jp}
\orcid{0000-0001-5045-341X}
\renewcommand{\shortauthors}{Yifan Li, Ryo Takahashi, et al.}

\begin{abstract}
Wireless mouse rings offer subtle, reliable pointing interactions for wearable computing platforms.
However, the small battery below 27 mAh in the miniature rings restricts the ring's continuous lifespan to just 1-10 hours, because even low-powered wireless communication such as BLE is power-consuming for ring's continuous use. 
The ring's short lifespan frequently disrupts users' mouse use with the need for frequent charging.
This paper presents picoRing \textit{mouse}, enabling a continuous ring-based mouse interaction with ultra-low-powered ring-to-wristband wireless connectivity. 
picoRing \textit{mouse} employs a coil-based impedance sensing named semi-passive inductive telemetry, allowing a wristband coil to capture a unique frequency response of a nearby ring coil via a sensitive inductive coupling between the coils.
The ring coil converts the corresponding user's mouse input into the unique frequency response via an up to 449 uW mouse-driven modulation system.
Therefore, the continuous use of picoRing \textit{mouse} can last approximately $600$~(8hrs use/day)-$1000$~(4hrs use/day) hours on a single charge of a $27$ mAh battery while supporting subtle thumb-to-index scrolling and pressing interactions in real-world wearable computing situations. 
\end{abstract}




\begin{teaserfigure}
  \includegraphics[width=\textwidth]{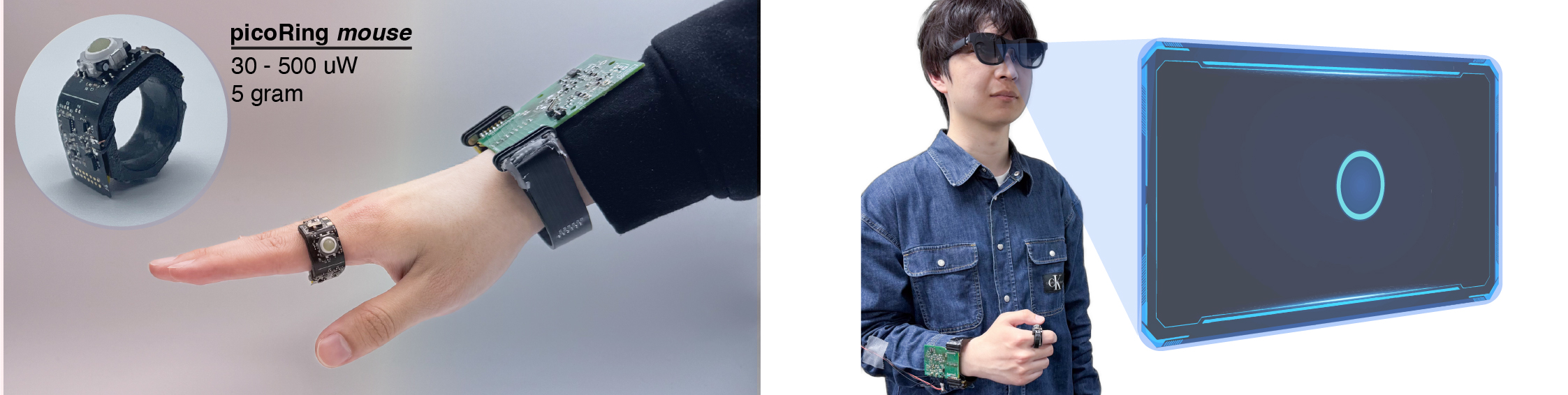}
  \caption{Overview of picoRing \textit{mouse}, enabling 30-500 uW-class ultra-low-power wireless ring mouse for ubiquitous finger input. The ring can potentially operate over a month on a single charge of a 27 mAh battery (\textcolor{blue}{\url{https://youtu.be/7RazVNMx0Ms})}.}
  \label{fig:overview}
  \Description{The image shows the picoRing mouse, an ultra-low-power wireless ring mouse. On the left, a close-up of the ring mouse is displayed, highlighting its compact design with electronic components. In the center, a person is wearing the ring mouse on their finger, with an additional electronic module on their wrist. On the right, a digital interface shows a blue ring on a screen, indicating interaction with the device. Text on the image reads "picoRing mouse, 30-500 uW, 5 gram."}
\end{teaserfigure}

\received{20 February 2007}
\received[revised]{12 March 2009}
\received[accepted]{5 June 2009}

\maketitle

\section{INTRODUCTION}

Mouse devices are basic tools for simple, fast pointing with computers~\cite{villar_mouse_2009}.
Integrating tinymouse functionality into wearable form factors such as wristbands, glasses and rings (\textit{e.g.,} Apple Watch, Galaxy Ring) promises always-available essential interactions with wearable computing platforms~\cite{xu_ao-finger_2023, kienzle_electroring_2021}.
Especially, the wireless ring-formed input devices worn on the index finger offer reliable detection of even subtle, privacy-preserved thumb-to-index finger inputs, unlike the wristband and glasses~\cite{takahashi_picoring_2024,kim_iris_2024}.
However, the physical constraint of the tiny ring structure requires the equipment of tens of \si{mAh} small batteries, challenging the continuous operation of tens of \si{\mW}-class power-consuming wireless communication modules. 
For example, prior wireless rings support below 1-10 hours of continuous wireless communication with Bluetooth Low Energy~(BLE), limiting the ring's communication usage to intermittent data transmission~\cite{zhou_one_2023, kim_iris_2024, shen_mousering_2024,jing_magic_2012}. 
Such an operation is suitable for periodic healthcare monitoring around the rings~(\textit{e.g.,} Oura Ring, Galaxy Ring), but not appropriate for the ring-based input interface that requires long-term, real-time communication with other wearables~(\textit{e.g.,} smartwatch, HMD).

This paper presents picoRing \textit{mouse}, enabling an ultra-low-power ring-shaped wireless tinymouse with a ring-to-wristband coil-based impedance sensing~(see \autoref{fig:overview}).
picoRing \textit{mouse} is inspired by a coil-based sensitive impedance sensing named passive inductive telemetry~(PIT) \cite{takahashi_picoring_2024, takahashi_telemetring_2020}.
Unlike long-range electromagnetic communication such as BLE and RF backscatter, PIT constructs a short-range but ultra-low-powered inductive link between a pair of a ring coil and a wristband coil.
Since the ring coil can send its sensor information to the wristband coil by simply modifying the inductive field generated from the wristband coil, the ring coil does not need active signal transmission with power-hungry communication modules.
With the combination of PIT with the ring-based mouse module, picoRing \textit{mouse} enables the ultra-low-powered tinymouse ring with maximum power consumption of approximately \SI{449}{\uW}, enabling continuous operation of approximately $7$ (24hrs use/day)-$44$ (4hrs use/day) days on a single charge of a \SI{27}{mAh} curved Lipo battery.
Furthermore, the shift from a previous full-passive ring design in original picoRing~\cite{takahashi_picoring_2024} to our semi-passive ring design by integrating an ultra-low-power MCU enables accurate detection of subtle mouse inputs without any physical noise issues (e.g. chattering) and demonstrates picoRing’s practical use in real-world scenarios. 
To facilitate replication and provide comprehensive implementation details, we have open-sourced the current picoRing \textit{mouse}:~\textcolor{blue}{\url{https://github.com/KawaharaLab/picoRing_mouse}}.

Our contribution is summarized as follows:
\begin{itemize}
    \item The design of an ultra-low-powered wireless communication between ring and wristband devices, inspired by sensitive PIT architecture~\cite{takahashi_picoring_2024}. 
    \item The demonstration and technical evaluation of \SI{500}{\uW}-class wireless ring mouse named picoRing \textit{mouse} toward ubiquitous thumb-to-index input.
\end{itemize}

\section{RELATED WORK}

\begin{figure}[t!]
  \centering
  \includegraphics[width=\columnwidth]{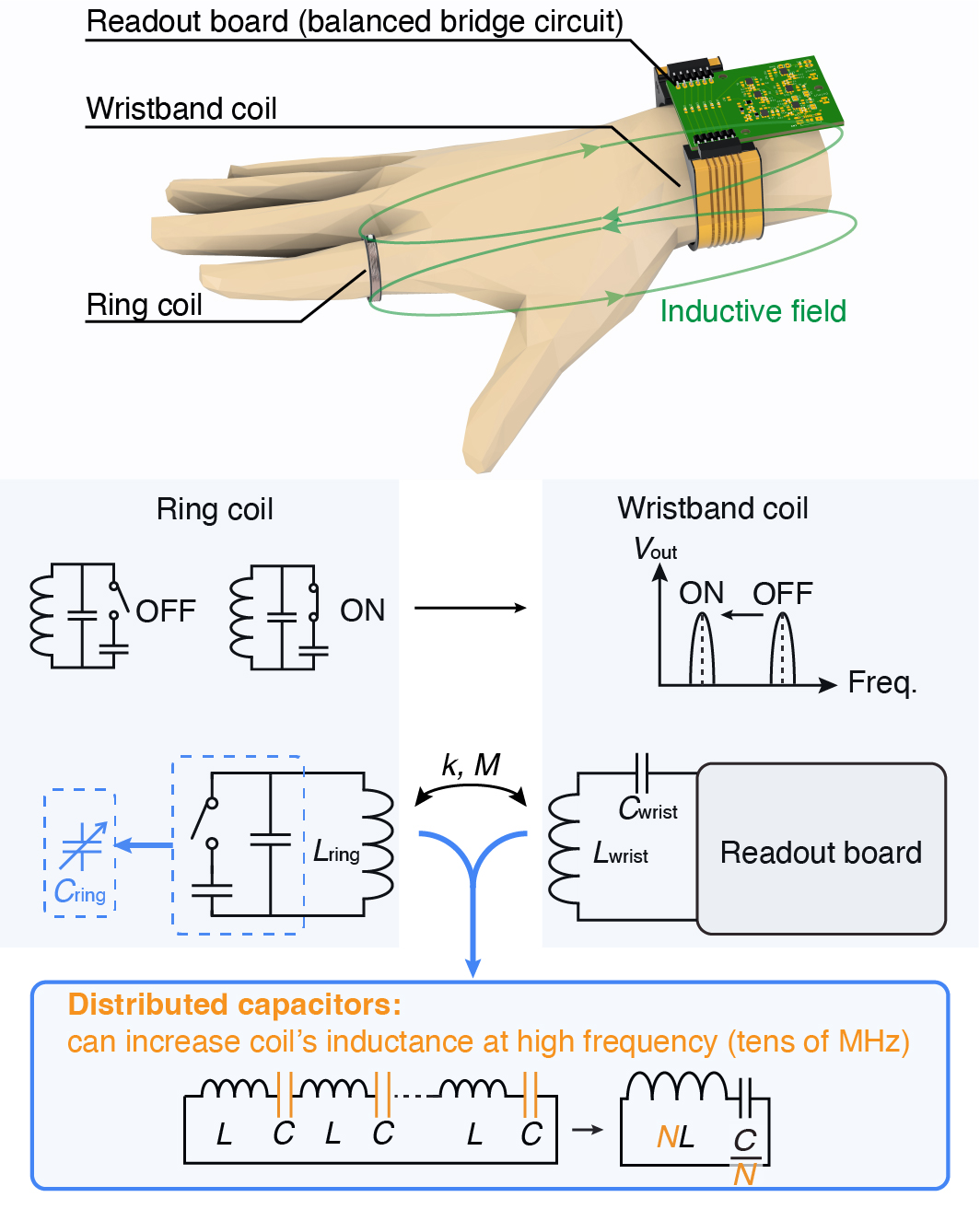}
  \caption{Schematic of passive inductive telemetry~(PIT) for ring-to-wristband low-powered wireless connection~\cite{takahashi_picoring_2024}. PIT allows the wristband coil to stably detect the passive response from the ring --the passive shift in the ring's resonant frequency-- through a ring-to-wristband inductive coupling.}
  \label{fig:pit}
  \Description{The schematic illustrates the passive inductive telemetry (PIT) system for a ring-to-wristband wireless connection in the picoRing device. At the top, a diagram shows a hand wearing a ring coil and a wristband coil connected to a readout board. Green arrows represent the inductive field between the ring and wristband. Below, circuit diagrams depict the ring coil's on/off states and its interaction with the wristband coil. A graph shows frequency shifts corresponding to the coil states. Annotations explain the role of distributed capacitors in increasing inductance at high frequencies.}
\end{figure}

\subsection{Wearable Mouse}

\begin{table*}[t!]
    \renewcommand{\arraystretch}{1.2}
    \centering
    \caption{Technical comparison of wireless ring-based input devices. To the best of our knowledge, picoRing \textit{mouse} is the first long-term wireless tinymouse ring achieving an ultra-low-powered operation below 1 mW.}
    \begin{tabular}{l|lllll}\toprule
        \textbf{Ring} & 
        \textbf{\myTableCell{Components}} & \textbf{\myTableCell{Wireless\\connection}} & \textbf{\myTableCell{Ring's maximum \\power \\(mW)}} & \textbf{\myTableCell{Available\\input interface}}  & \textbf{\myTableCell{User and session\\independence}}\\\hline\hline
        Galaxy Ring& Ring & BLE & - & Finger tap & \pmb{\checkmark}\\ 
        \rowcolor{myblue!20}ElectroRing~\cite{kienzle_electroring_2021}& Ring & BLE & $220$ & Finger press & \pmb{\checkmark}\\ 
        Ring-a-Pose~\cite{yu_ring--pose_2024}& Ring & BLE & $148$ & Finger pose & \pmb{\checkmark}\\ 
        \rowcolor{myblue!20}OmniRing~\cite{zhou_one_2023}& 1-5 rings & BLE & $40$/ring & Finger pose & \pmb{\checkmark}\\ 
        IRIS~\cite{kim_iris_2024}& Ring & BLE & $26$ & Vision & \pmb{\checkmark}\\ 
        \rowcolor{myblue!20}MouseRing~\cite{kim_iris_2024}& Ring & BLE & $-10$ & Finger slide & \pmb{\checkmark}\\ 
        FingeRing~\cite{fukumoto_fingering_1994}& 5 rings \& wristband & \myTableCell{Intrabody\\communication} & \textbf{$\mathbf{1.8}$/ring} & Finger tap & \myTableCell{$\triangle$\\Unstable link \\under RF noises}\\
        \rowcolor{myblue!20}Nenya~\cite{ashbrook_nenya_2011}& Ring \& wristband & \myTableCell{Static\\magnetic field} & \textbf{0} & Finger scroll & \myTableCell{$\times$\\Unstable link\\under magnetic noises}\\
        AuraRing~\cite{parizi_auraring_2019}& Ring \& wristband & \myTableCell{Near-field\\inductive coupling} & $\mathbf{2.3}$ & Finger pose & \pmb{\checkmark}\\
        \rowcolor{myblue!20}TelemetRing~\cite{takahashi_telemetring_2020}& 5 rings \& wristband & PIT & \textbf{$\mathbf{0}$/ring} & Finger tap & \pmb{\checkmark}\\
        picoRing~\cite{takahashi_picoring_2024} & Ring \& wristband & PIT & $\mathbf{0}$& \myTableCell{Finger press \textbf{or} scroll}  & \pmb{\checkmark}\\
        \rowcolor{myblue!20}\textbf{picoRing \textit{mouse}} & Ring \& wristband & semi-PIT & $\mathbf{0.45}$ & \myTableCell{Finger press \textbf{and} scroll} & \pmb{\checkmark}\\
        \bottomrule
    \end{tabular}
    \label{tab:comparison}
    \Description{The table compares technical specifications of various wireless ring-based input devices. Columns include the ring name, components, wireless connection type, ring's maximum power consumption in mW, available input interfaces, and user/session independence. The picoRing mouse is highlighted as the first device achieving ultra-low power consumption below 1 mW, specifically 0.45 mW, using a semi-PIT connection. Other devices listed include ElectroRing, Ring-a-Pose, OmniRing, and more, with varying power consumption and input methods.}
\end{table*}

Wearable mouse has emerged as seamless interaction with wearable computing devices~\cite{clarke_spectrum_1994,rekimoto_gesturewrist_2001,fukumoto_body_1997}. 
Specifically, the smart accessories such as rings or wristbands allows users to control and interact with technology through subtle finger and hand movements~\cite{jiang_emerging_2022}.
Wristbands have been explored for detecting finger movements using various technologies, including force measurement~\cite{dementyev_wristflex_2014}, electromagnetic waves~\cite{kim_etherpose_2022}, inertial sensing~\cite{laput_viband_2016}, capacitive coupling~\cite{zhang_advancing_2016}, and electromyography (EMG)~\cite{saponas_enabling_2009}.
While these methods can detect gestures, they face challenges in accurately capturing subtle finger motions due to the distance from the fingers. 
Machine learning and sensor fusion techniques have been employed to improve gesture recognition accuracy, though they often require dynamic motions and frequent calibration~\cite{xu_ao-finger_2023}.
In contrast, rings provide a closer proximity to the fingers, allowing for more precise detection of subtle movements even with simple sensors and algorithms~\cite{takahashi_picoring_2024,kienzle_electroring_2021}. 
This close-range sensing enables stable interaction without causing user fatigue, making them suitable for everyday use.

\subsection{Wireless Ring}
Ring-formed mouse devices have long been investigated within the HCI community.
The exploration of the previous mouse ring can be divided into two categories: i) ring-based sensing techniques such as computer vision~\cite{sun_thumbtrak_2021, chan_cyclopsring_2015, kienzle_lightring_2014}, inertial sensors~\cite{shen_mousering_2024, liang_dualring_2021}, ultrasound~\cite{yu_ring--pose_2024,zhang_fingersound_2017}, and electromagnetic wave~\cite{chen_efring_2022, waghmare_z-ring_2023} and ii) low-powered wireless ring design using RF backscatter~\cite{liu_ambient_2013,ensworth_ble-backscatter_2017,bainbridge_wireless_2011}, near-field communication~(NFC)~\cite{zhan_flexible_2024,lee_nfcstack_2022,liang_nfcsense_2021}, intrabody communication~\cite{fukumoto_body_1997,varga_designing_2018}, and magnet-based tracking~\cite{ashbrook_nenya_2011}.
The ring-based finger sensing demonstrates high-fidelity microgesture recognition compared to the wristband- and glasses-based finger sensing~\cite{kienzle_electroring_2021}.
However, the long-term operation of the low-powered wireless rings is still challenging~\cite{wagih_microwave-enabled_2023}.
Specifically, \si{\uW}-class RF backscatter and magnet-mounted ring need a large antenna and a bulky magnet, which is hard to fit in the small ring~\cite{naderiparizi_towards_2018,ashbrook_nenya_2011}.
Moreover, the magnet-based tracking via exposed static magnetic field is easily influenced by the surrounding magnetic noises such as electrical appliances, impairing the tracking accuracy of the magnet position or orientation~\cite{wang_automatic_2022}.
\si{\mW}-class NFC supports small tag design, but the communication range is a few centimeter-scale~\cite{gummeson_energy_2014, takahashi_full-body_2025,zhang_nfcapsule_2022}; such a short-range link cannot construct ring-to-wristband communication.
Intrabody communication~\cite{fukumoto_body_1997,varga_designing_2018}, which uses capacitive body as a signal path, can transmit signal with \si{\uW}-class ultra-low-power throughout the body~\cite{li_body-coupled_2021}, but both the RF presence of other electronic devices and the daily items such as wooden desk and wall can impair the capacitive signal path, causing signal degradation, reduced transmission range, or even data loss~\cite{kienzle_electroring_2021}.

\subsection{Passive Inductive Telemetry}

\begin{figure*}[t]
  \centering
  \includegraphics[width=1.0\textwidth]{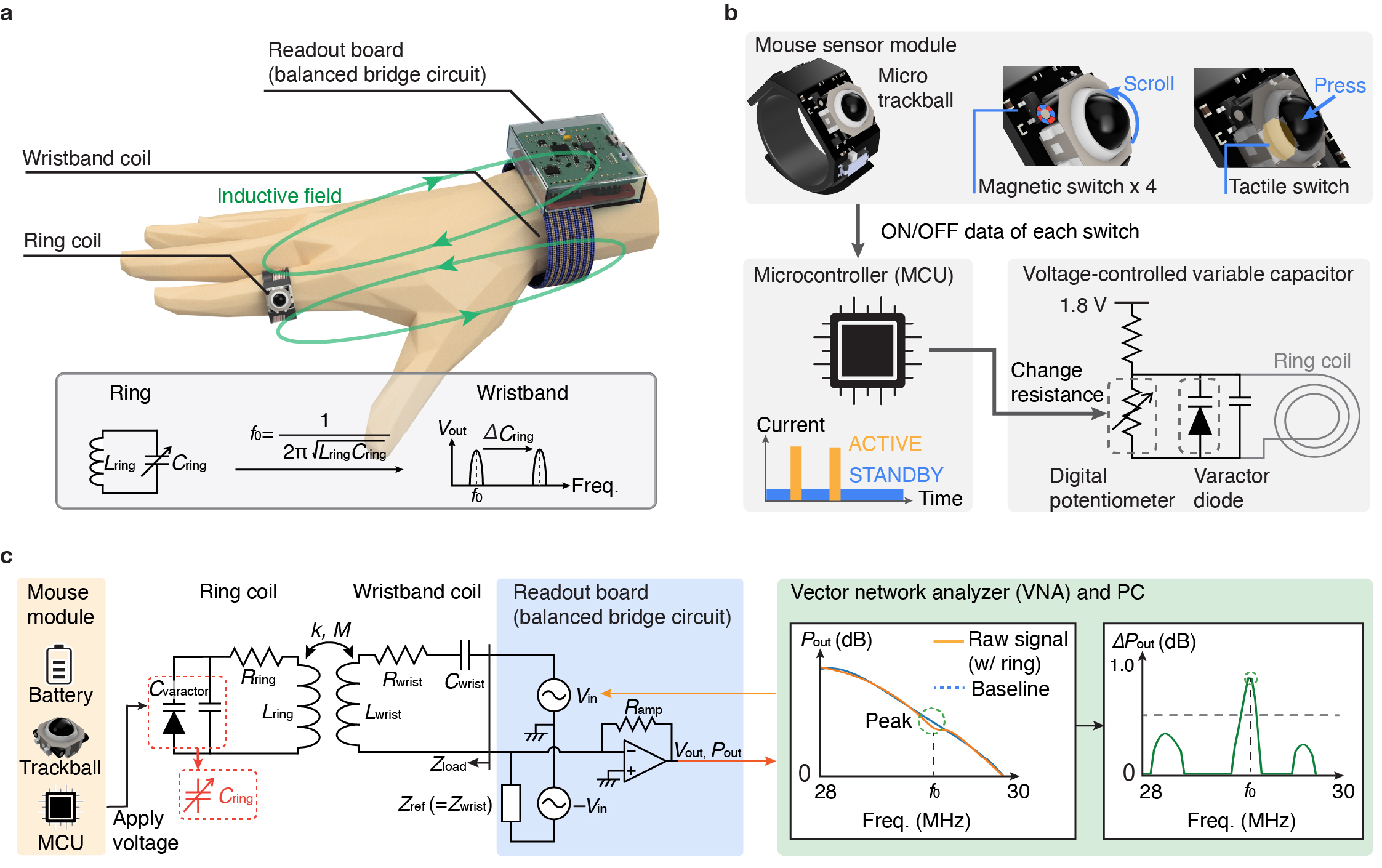}
  \caption{Design overview of picoRing \textit{mouse}. (a) Schematic of picoRing \textit{mouse}, consisting of a pair of a ring coil and a wristband coil connected to a readout board (impedance measurement circuit). (b) Illustration of working principle of the ring coil to convert the mouse input into unique shift of the ring's resonant frequency. (c) Circuit diagram of picoRing \textit{mouse}.}
  \label{fig:design}
  \Description{The design overview of the picoRing mouse is shown in three parts. (a) A schematic depicts the ring coil and wristband coil connected to a readout board, illustrating the inductive field and resonant frequency shift. (b) Illustrations show the mouse sensor module with magnetic and tactile switches, connected to a microcontroller (MCU). The diagram explains how input is converted into electrical signals, highlighting active and standby current states. (c) The circuit diagram displays components like the ring coil, wristband coil, and readout board, detailing the vector network analysis process for frequency measurement.}
\end{figure*}

picoRing~\cite{takahashi_picoring_2024}, which is most similar to this paper, extends the coil-based communication distance by using PIT~\cite{takahashi_twin_2021,takahashi_telemetring_2020,li_demo_2025}.
PIT consists of a pair of a fully-passive ring coil and a wristband coil, allowing the wristband to detect the passive response of the ring coil~(see \autoref{fig:pit}).
Specifically, ring coil, which consists of a resonant $LC$ tank connected to a physical switch, converts the thumb-to-switch inputs into the change of the ring’s resonant frequency, defined as $f=1/(2\pi\sqrt{LC})$, without any active electrical components; ON/OFF of the switch connected to the chip capacitor allows the passive change of the total capacitance of the ring coil.
Because the ring's resonant change causes the impedance change in the wristband coil through ring-to-wristband inductive coupling, the wristband coil can recognize the thumb-to-switch input by detecting the frequency peak in its impedance characteristics, appeared around the ring's resonant frequency.                                      
Because the ring away from the wristband inevitably couples with the wristband through a significantly weak inductive coupling, prior picoRing~\cite{takahashi_picoring_2024} has proposed the sensitive PIT by combining a sensitive coil with distributed capacitors~\cite{cook_large-inductance_1982} with a sensitive impedance measurement circuit named balanced bridge circuit~\cite{takahashi_telemetring_2020}.
The sensitive coil, which has a large turn number or large inductance at a high resonant frequency (\textit{i.e.,} tens of \si{\MHz}) by inserting multiple chip capacitors into a single long coil, can increase the passive response from the ring coil.
Furthermore, the bridge circuit, which uses a differential circuit structure and impedance balancing technique, can be sensitive to the small impedance change of the wristband coil. 
With these configuration, picoRing constructs stable ring-to-wristband wireless link despite proximity to metallic items and electrical appliances.
For more detail, please refer to~\cite{takahashi_picoring_2024}.

While picoRing employs the passive mechanical switch as the battery-free variable capacitor, the ring without any digital computational components, e.g., microcontroller~(MCU) and analog-to-digital converter, is challenging to support mouse-like multi-modal inputs.
This is because the ring-sized input sensors available in the commence needs the battery-assisted electrical sensors; the miniaturization of multi-modal mechanical switch is hard due to the spatial challenges. 
Therefore, prior picoRing~\cite{takahashi_picoring_2024} requires the users to change the different types of rings~(\textit{e.g.,} picoRing \textit{press}, \textit{slide}, \textit{joystick}, and \textit{scroll}) according to the target finger inputs.
By contrast, picoRing \textit{mouse} can support multi-modal thumb-to-index inputs with a single ring by integrating low-powered, miniaturized mouse sensor and signal encoding circuit into the prior picoRing system.
\autoref{tab:comparison} shows the technical comparison among ring-formed mouse devices.
\textbf{To the best of our knowledge, picoRing \textit{mouse} is the first to demonstrate the ultra-low-powered wireless mouse ring below $\mathbf{500}$ uW for long-term ubiquitous finger input interface.}

\section{SYSTEM DESIGN}

picoRing \textit{mouse} consists of two main components~(see ~\autoref{fig:design}a): 1) a ring coil with a mouse module, which changes its frequency response based on the thumb-to-index inputs, and 2) a wristband coil that detects the unique peak in the frequency response generated by the ring's resonance.
The working principle of picoRing \textit{mouse} is as follows:
First, the wristband coil generates a weak inductive field to couple with the nearby ring coil.
When the user scrolls or presses the mouse module mounted on the ring, the MCU connected to the mouse module detects either the scrolling or pressing via the magnet or tactile switch, transforming the mouse input into a unique frequency response~(\textit{i.e.,} the change in the ring's resonant frequency) via a voltage-controlled variable capacitor~(see \autoref{fig:design}b).
Owing to the inductive coupling, the wristband coil connected to a sensitive impedance measurement module obtains the frequency response.
Since the ring's frequency response appears as a sharp frequency peak at the ring's resonant frequency in the frequency characteristics of the wristband's impedance, the wristband coil enables to recognize the mouse input~(\textit{i.e.,} pressing and 2D scrolling) as a unique peak output.

\subsection{Ring-to-wristband Wireless Communication}
\label{sec:semi-PIT}

%TODO
picoRing \textit{mouse} extends the PIT system in a semi-passive manner.
While the prior PIT features the full-passive design of the ring coil by modifying the ring-to-wristband inductive field with a full-passive modulation circuit composed of tactile switches.
In contrast, picoRing \textit{mouse} employs the semi-passive modulation approach, which modifies the inductive field with a low-powered digital modulation circuit composed of a battery-assisted voltage-controlled variable capacitor.
Note that the semi-passive modulation approach does not use the active signal transmission circuit.
Unlike full-passive modulation, the semi-passive modulation can transmits the sensor information as digital signals.
Specifically, picoRing \textit{mouse} uses a simple frequency-shift keying, which encodes the multi-modal mouse input into a corresponding frequency shift based on the voltage-controlled variable capacitor.

Let us explain how the ring coil can influence the wristband coil in the frequency-shifting manner.
When the ring coil resonates at $f_0$ inductively couples with the wristband coil, the input impedance of the wristband coil, $Z_{\rm load}$, can be expressed as follows:
\begin{align}
    Z_{\rm load}(\omega) &= Z_{\rm wrist}(\omega) + \Delta Z_{\rm wrist}(\omega)\label{eq:z_load}\\
    \Delta Z_{\rm wrist}(\omega) &= \cfrac{(\omega M)^2}{Z_{\rm ring}(\omega)}  \label{eq:delta_z_wrist}\\
    Z_{\rm wrist}(\omega) &= R_{\rm wrist} + j\left(\omega L_{\rm wrist} - \cfrac{1}{\omega C_{\rm wrist}}\right)\\
    Z_{\rm ring}(\omega) &= R_{\rm ring} + j\left(\omega L_{\rm ring} - \cfrac{1}{\omega C_{\rm ring}}\right)
\end{align}
where $Z_{\rm wrist}$ and $Z_{\rm ring}$ are the total impedance of the wristband and ring coils, respectively, $\omega~ (=2\pi f)$ is the angular frequency, $M$ is the mutual inductance between the ring and wristband coils.
$Z_{\rm ring}$ can be simplified as $R_{\rm ring}$ at $f_0$ because $\omega_0 L_{\rm ring} - 1/(\omega_0 C_{\rm ring})$ is $0$ in the resonance state.
As a result, \autoref{eq:delta_z_wrist} can be also simplified at $\omega_0 = 2\pi f_0$ as follows:
\begin{align}
    \Delta Z_{\rm wrist}(\omega_0) =\cfrac{(\omega_0 M)^2}{R_{\rm ring}} \label{eq:z_load_max}
\end{align}
\autoref{eq:z_load_max} indicates $\Delta Z_{\rm wrist}$ increases significantly around $\omega_0$, resulting in the sharp peak in the frequency characteristics of $\Delta Z_{\rm wrist}$ around $\omega_0$.
However, similar to picoRing~\cite{takahashi_picoring_2024}, the extremely weak inductive coupling makes $\Delta Z_{\rm wrist}$ too small to detect through a standard impedance measurement circuit.

To detect $\Delta Z_{\rm wrist}$, or small variations in $Z_{\rm load}$, the balanced bridge circuit is useful~\cite{takahashi_picoring_2024,takahashi_telemetring_2020}.
The balanced bridge circuit, shown in \autoref{fig:design}c, can be sensitive to only the small impedance change, by an impedance balancing process between the wristband coil and embedded chip components on the bridge circuit~($Z_{\rm wrist} = Z_{\rm ref}$).
With this impedance balance, the output of the bridge circuit, $V_{\rm out}$ and $P_{\rm out}$, at $\omega_0$ can be expressed with $Z_{\rm wrist} \gg \Delta Z_{\rm wrist}$, \autoref{eq:z_load}, and \autoref{eq:delta_z_wrist}:
\begin{align}
    V_{\rm out}(\omega) &=  -R_{\rm amp} \left(\cfrac{V_{\rm in}}{Z_{\rm load}} - \cfrac{V_{\rm in}}{Z_{\rm ref}}\right)\\
    &\approx  R_{\rm amp} \cfrac{\Delta Z_{\rm wrist}(\omega)}{Z_{\rm ref}^2}V_{\rm in} \\
    &\downarrow \mbox{convert $V_{\rm out}$ to $P_{\rm out}$~(\si{\dB})}\notag\\
    \Delta P_{\rm out}(\omega) (\si{\dB}) &\approx
    \begin{cases}
    0~~~~~(\omega \neq \omega_0~\mbox{or w/o ring coil})\\
    \Delta Z_{\rm wrist}(\omega_0)~(\si{\dB})~~~~~(\omega \approx \omega_0)
    \end{cases}\label{eq:bridge_output}
\end{align}
where $V_{\rm in}$ is the input voltage of the bridge circuit, $R_{\rm amp}$ is the amplifier factor.
Note that $Z_{\rm ring}~(\omega) \gg (\omega M)^2$ is valid under condition $\omega \neq \omega_0$ because $Z_{\rm ring}$ has the large imaginary part at non-resonant frequency and $M$ is so small in the situation where the ring coil is away from the wristband coil.
\autoref{eq:bridge_output} indicates that the wristband coil can reliably detect the ring's resonant frequency through output~($P_{\rm out}(\si{\dB})$) of the bridge circuit.
Therefore, the ring coil, which changes its resonant frequency with the variable capacitor, can send its mouse sensor information to the wristband coil in the frequency-keying way.
Moreover, the semi-passive signal transmission of the ring coil, which modifies the inductive field from the wristband coil in the load-modulation way, can save power for signal transmission unlike the active signal transmission.

\subsection{Ring}
\label{sec:ring}

\begin{figure}[t!]
  \centering
  \includegraphics[width=1\columnwidth]{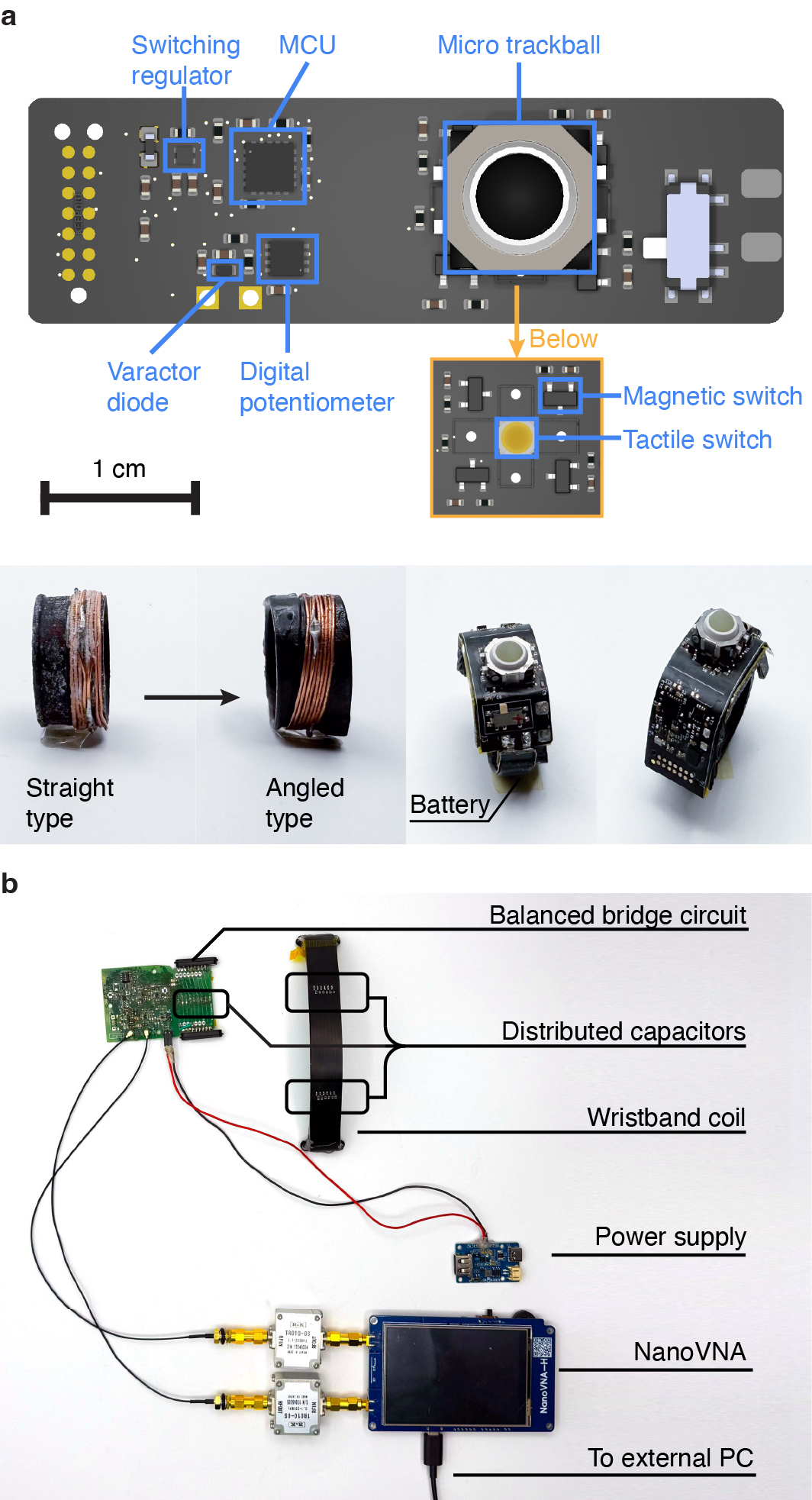}
  \caption{Prototype photograph of our (a) ring coil with a $\mathbf{45\times14}$ mm flexible PCBs including a micro trackball, mouse sensor module, MCU, varactor diode, digital potentiometer, and switching regulator, and (b) wristband coil implemented flexible PCBs. The flexible PCBs are mounted to the 3D-printed wristband, and the wristband coil is connected to the readout board, composed of the balanced bridge circuit, NanoVNA, and external PC.}
  \label{fig:implementation}
  \Description{The photograph shows the prototype of the picoRing mouse. (a) Displays the ring coil on a 45 × 14 mm flexible PCB with labeled components: magnetic mouse, mouse sensor module, MCU, varactor diode, digital potentiometer, and switching regulator. Below, images show the ring coil in straight and angled types, mounted on a 3D-printed structure. (b) Shows the wristband coil on flexible PCBs connected to a readout board, including a balanced bridge circuit, NanoVNA, and external PC. Components like distributed capacitors and power supply are labeled.}
\end{figure}

\begin{figure*}[ht!]
  \centering
  \includegraphics[width=1\textwidth]{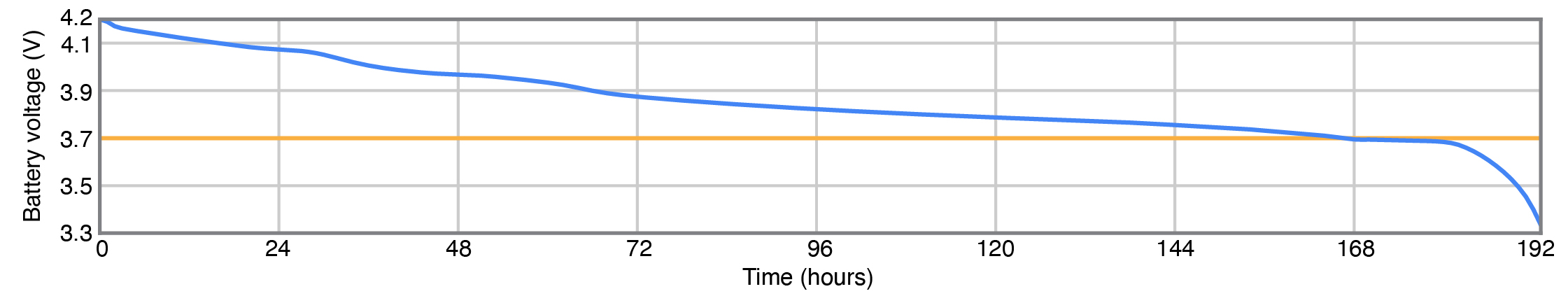}
  \caption{Data plot of battery voltage over the operation time of our wireless ring.}
  \label{fig:eva_operation_hours}
  \Description{The graph shows battery voltage (V) over time (hours) for the wireless ring. The blue line represents battery voltage decreasing from 4.2V to just below 3.7V over 192 hours. A horizontal orange line at 3.7V indicates a reference voltage level. The plot illustrates a gradual decline in voltage over time, with a sharper drop near the end.}
\end{figure*}

The ring coil consists of an $8$-turned resonant coil with distributed capacitors and the mouse module equipped with a small trackball~(EVQWJN007, Panasonic) supporting up, down, right, left scrolling and pressing interactions. 
When users scroll or press the trackball in the mouse module, the trackball either i) converts the scroll rotation of the trackball into rotation of cylindrical magnets with alternating polarity distribution or ii) turns on the tactile switch reacting to the user press of the trackball (see \autoref{fig:design}b).
The magnetic switch (CT8132BL, Allegro MicroSystems) below each magnet sends the magnet rotation data to an ultra-low-power MCU (STM32L011F4U6, STM) along with the ON/OFF data of the tactile switch, and then, according to the input type, the MCU applies a different corresponding voltage to a varactor diode~(SMV1253, Skyworks) by a digital-potentiometer-based voltage divider circuit~(AD5160, Analog Devices). 
The varactor diode, which varies its capacitance from \SI{27}{\pF} to \SI{69}{\pF} based on the applied voltage, can change the $C_{\rm ring}$ by being connected in parallel to one of the distributed capacitors in the ring coil.

The ring coil provides five physical inputs: none, scroll down, up, left, right, and press.
The resonant frequency for each input is set to \SI{27.32}{\MHz}, \SI{27.46}{\MHz}, \SI{27.60}{\MHz}, \SI{27.83}{\MHz}, \SI{28.23}{\MHz}, and \SI{28.47}{\MHz}, respectively, within the impedance-balanced frequency band.
The inductance~($L_{\rm ring}$) and resistance~($R_{\rm ring}$) of the ring coil at \SI{28.2}{\MHz} are \SI{2.6}{\uH} and \SI{3.5}{\ohm}, respectively.
Note that the ring coil is tilted with approximately $20^{\circ}$, mitigating the misalignment between the ring and wristband coils when the hand is grasped for thumb-to-index input.
Scrolling diagonally results in the simultaneous scrolling of adjacent horizontal and vertical magnets.
For instance, when scrolling diagonally to the upper right, both the up and right magnets move concurrently.
The MCU decodes this diagonal scrolling into vertical and horizontal scrolling in turn.

The ring coil has two types of operation modes: ACTIVE mode and STANDBY mode.
In the ACTIVE mode, the ring coil continuously streams data to the wristband coil with $200$ fps.
By contrast, in the STANDBY mode, the timer in the MCU periodically awakens to check the receiving data from the connected magnet switch.
picoRing basically waits in the STANDBY mode and transits to ACTIVE mode when the user inputs to the mouse module. 
When there is no signal change in the magnet switch for \SI{30}{\s}, picoRing backs to the STANDBY mode.
Note that the MCU clock in the ACTIVE and STANDBY modes is set to \SI{524}{\kHz} and \SI{32}{\kHz}, respectively.

\subsection{Wristband}
\label{sec:wrist}

The wristband coil consists of a $6$-turned flexible resonant coil mounted on a 3D-printed flexible wristband and the readout board connected to a mobile vector network analyzer~(VNA) (NanoVNA-H4). 
To recognize the ring's mouse state, the wristband coil detects the ring's frequency peak~($f_0$) by monitoring the frequency sweep signal ranging from \SI{27.0}{\MHz} to \SI{28.5}{\MHz} with an input power of \SI{0.2}{\mW}, the frequency bandwidth of \SI{4}{\kHz}, the sampling number of $101$, and the sampling rate of $20$ fps.
This setup for fast VNA readout enables the reliable detection of subtle and fast finger scrolling but, the frequency peak sometimes shifts because the large frequency step resolution of the NanoVNA without noise reduction sometimes fluctuates the frequency peak data.
To address this, we’ve designed the ring’s six resonant frequencies with a enough frequency space of $0.2-0.4$~\si{\MHz}, as described in \S~\ref{sec:ring}.

The wristband created by FDM 3D-printing of TPU filament is designed in the ellipse shape with \SI{6.2}{\cm} width, \SI{5.2}{\cm} height, and \SI{1.9}{\cm} length to fit in the middle-sized adult hands.
The wristband coil is connected to the readout board via two $6$ pin magnetic connectors (DIY Magnetic Connector, Adfruit). 
The reader board consists of the bridge circuit composed of six amplifiers (LTC6228, Analog Devices), magnetic connectors, and an external PC (Macbook Air).
The NanoVNA sends the power spectrum of the frequency response to the connected external PC, which analyzes the frequency peak in the power spectrum, similar to the same peak detection algorithm as picoRing~\cite{takahashi_picoring_2024}.
In total, the resonant frequency, inductance~($L_{\rm wrist}$) and resistance~($R_{\rm wrist}$) of the wristband coil connected with eighteen \SI{150}{\pF} distributed capacitors and a \SI{47}{\ohm} resistor in series are \SI{27}{\MHz}, \SI{4.2}{\uH}, and \SI{49}{\ohm}, respectively, and the measured power consumption of the bridge circuit and NanoVNA are \SI{0.33}{\W}~($=\SI{3.3}{\V}\times\SI{0.1}{\A}$) and \SI{1.0}{\W}~($=\SI{5}{\V}\times\SI{0.2}{\A}$), respectively.
The current wristband coil created by the flexible PCBs and TPU-based band would have the challenge of wearability and stretchability.
The use of stretchable textile coil~\cite{kanada_joint-repositionable_2025,takahashi_twin_2021} is promising for high biocompatibility for the skin.

\section{TECHNICAL EVALUATION}

\begin{table}[t!]
    \centering
    \caption{Power consumption of the ring coil hardware.}
    \begin{tabular}{lll}\toprule
        \textbf{Component} & \textbf{\myTableCell{STANDBY\\(\si{\uW})}}  & \textbf{\myTableCell{ACTIVE\\(\si{\uW})}}\\\hline
        MCU (STM32L011F4U6) & $1.8$ & $252$ \\
        Mouse sensor module (4 $\times$ CT8132) & $7.9$ & $7.9$ \\
        Digital voltage divider (MAX5394) & $1.8$ & $36$ \\
        Power management (TPS62843) & $0.63$ & $0.63$ \\\hline
        \textbf{Estimated total} & $>12$ & $>296$\\
        \rowcolor{myblue!20}\textbf{Measured total@4.2~V} & $\mathbf{25}$ & $\mathbf{449}$\\\bottomrule
    \end{tabular}
    \label{tab:power}
    \Description{The table lists power consumption of ring coil hardware components in standby and active modes, measured in microwatts (µW). Components include the MCU (STM32L011F4U6), mouse sensor module (4 × CT8132), digital voltage divider (MAX5394), and power management (TPS62843). Standby consumption ranges from 0.63 to 7.9 µW, while active consumption ranges from 0.63 to 252 µW. Estimated total is over 12 µW in standby and over 296 µW in active. Measured total at 4.2V is 25 µW in standby and 449 µW in active.}
\end{table}

\begin{table}[t!]
    \centering
    \caption{Estimation of battery lifespan of the ring coil driven by 20 mAh or 27 mAh batteries, respectively.}
    \begin{tabular}{lll}\toprule
        \textbf{\myTableCell{Operation time of\\ACTIVE mode\\(hrs/day)}}  & \textbf{\myTableCell{Battery lifespan\\w/ 20 mAh\\(hrs)}}  & \textbf{\myTableCell{Battery lifespan\\w/ 27 mAh\\(hrs)}}\\\hline
        $24$ & $167^{1}$  & $225$\\
        \rowcolor{myblue!20}$8$ & $451$ & $609$ \\
        \rowcolor{myblue!20}$4$ & $784$  & $1058$\\\bottomrule
    \end{tabular}\\
    $^{1}$: We measured the actual battery lifespan for this case.
    \label{tab:battery_lifespan}
    \Description{The table estimates battery lifespan of the ring coil using 20 mAh and 27 mAh batteries. For 24 hours of active mode per day, the lifespan is 167 hours for 20 mAh and 225 hours for 27 mAh. For 8 hours per day, it's 451 hours for 20 mAh and 609 hours for 27 mAh. For 4 hours per day, it's 784 hours for 20 mAh and 1058 hours for 27 mAh. A footnote indicates the 167-hour lifespan for 20 mAh was measured.}
\end{table}

\subsection{Operation Time of picoRing \textit{mouse}}

First, we measured how long picoRing \textit{mouse} can operate.
To calculate the expected battery life during usage, we first connected the battery terminals of picoRing \textit{mouse} to a digital multimeter~(34410A, Keysight).
The measured power consumption at \SI{4.2}{\V} during ACTIVE and STANDBY mode is measured to be approximately \SI{449}{\uW} ($=$\SI{4.2}{\V}$\times$\SI{107}{\uA}) and \SI{35}{\uW}~($=$\SI{4.2}{\V}$\times$\SI{8.4}{\uA}), respectively.
Then, we estimated the battery lifespan hours for three types of different ACTIVE operation time of picoRing \textit{mouse}.
For reference, we measured the battery lifespan hours when picoRing \textit{mouse} operates continuously in ACTIVE mode.
The battery lifespan is calculated from when the LiPo battery is fully charged at \SI{4.2}{\V} until it drops below \SI{3.7}{\V}~(see \autoref{fig:eva_operation_hours}).
To maintain the ACTIVE mode, the MCU is instructed to keep the digital voltage divider ON, i.e., shifting up the resonant frequency of the ring-shaped coil from \SI{27}{\MHz}~(\textit{i.e.}, the digital voltage divider is OFF) to \SI{29}{\MHz}.
Note that the magnetic scroll detector is also kept running continuously.
The measured time is approximately $167$ hours, i.e., 7 days with the fully-charged \SI{20}{mAh} Lipo battery~(UFX150732).

Based on the result of power consumption and battery lifespan in continuous ACTIVE mode, \autoref{tab:battery_lifespan} shows the expected battery lifespan for the different operation hours (\textit{i.e.,} 4, 8, or 24 hours per day).
The lifespan hour is estimated based on two types of battery capacity: 20 mAh and 27 mAh, both of which is widely used in commercially-available smart ring products.
Based on the practical usage time of a mouse without causing tendinitis, which is approximately 4-8 hours per day~\cite{mikkelsen_validity_2007,gerr_prospective_2002,wolff_comparing_2018}, picoRing \textit{mouse} has the potential to operate for approximately $451$ (8 hrs/day)-$1058$ (4 hrs/day) hours, or over a few weeks or one month, on a single charge of $20-27$ mAh battery.

\begin{figure*}[ht!]
  \centering
  \includegraphics[width=1.0\textwidth]{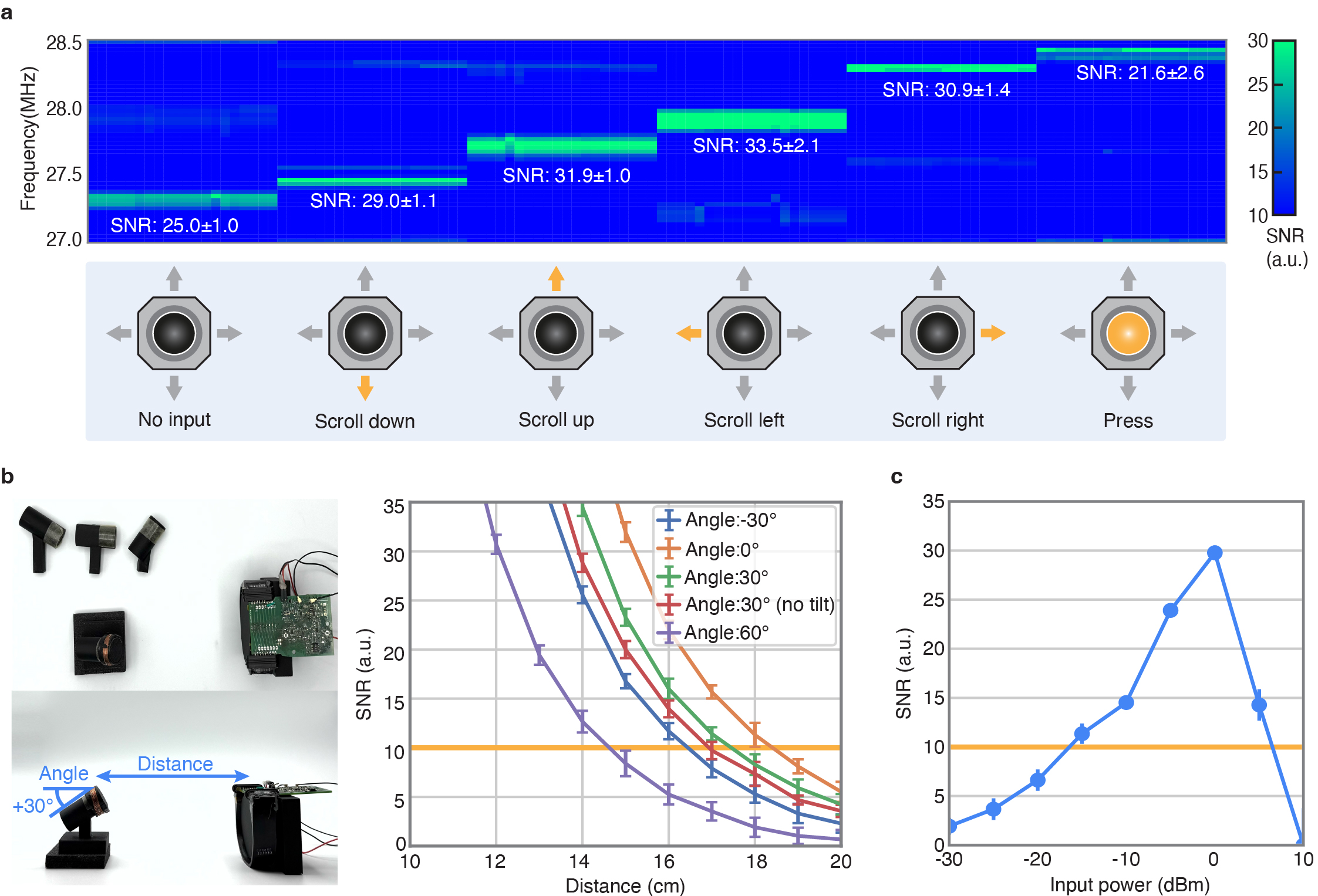}
  \caption{Signal-to-noise ratio~(SNR) of picoRing \textit{mouse} in controlled environments. (a) Time-series and average SNR for scrolling and pressing interactions for three users. (b) SNR for the distance and angle between the ring and the wristband coils. (c) SNR for the input power of the wristband coil. }
  \label{fig:snr_in_controlled}
  \Description{Figure 6 showcases the Signal-to-Noise Ratio (SNR) of the picoRing mouse in controlled environments.
(a) A time-series graph illustrates the SNR for scrolling and pressing interactions among three users, with frequency on the y-axis and SNR values annotated for each interaction type: no input, scroll down, scroll up, scroll left, scroll right, and press. (b) Two images show different angles and distances between the ring and wristband coils, accompanied by a graph plotting SNR against distance at various angles (-30°, 0°, 30°, 30° no tilt, and 60°). (c) A graph displays SNR as a function of input power to the wristband coil, with input power on the x-axis and SNR on the y-axis, highlighting a peak SNR at a specific input power level.}
\end{figure*}

\begin{figure*}[ht!]
  \centering
  \includegraphics[width=1.0\textwidth]{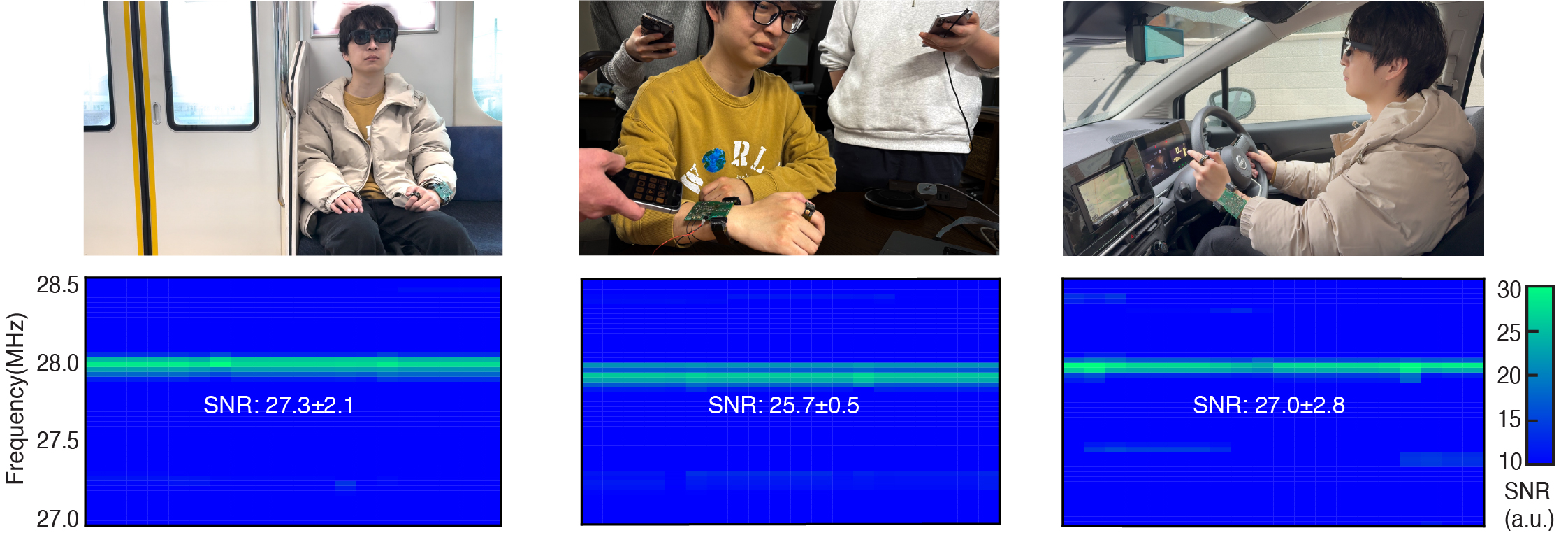}
  \caption{Signal-to-noise ratio~(SNR) of picoRing \textit{mouse} against the three types of electromagnetically-noisy sessions in real-world wearable computing.}
  \label{fig:snr_in_real}
  \Description{Figure 7 displays the Signal-to-Noise Ratio (SNR) of the picoRing mouse during three electromagnetically noisy scenarios in real-world wearable computing. The top row shows images of each scenario: a person sitting on public transport, a group setting with multiple people using electronic devices, and a person driving a car while using the device. The bottom row contains corresponding SNR graphs for each scenario, with frequency on the y-axis and annotated SNR values of 27.3±2.1, 25.7±0.5, and 27.0±2.8. A color bar on the right indicates the SNR scale from 10 to 30.}
\end{figure*}

\begin{figure*}[ht!]
  \centering
  \includegraphics[width=1.0\textwidth]{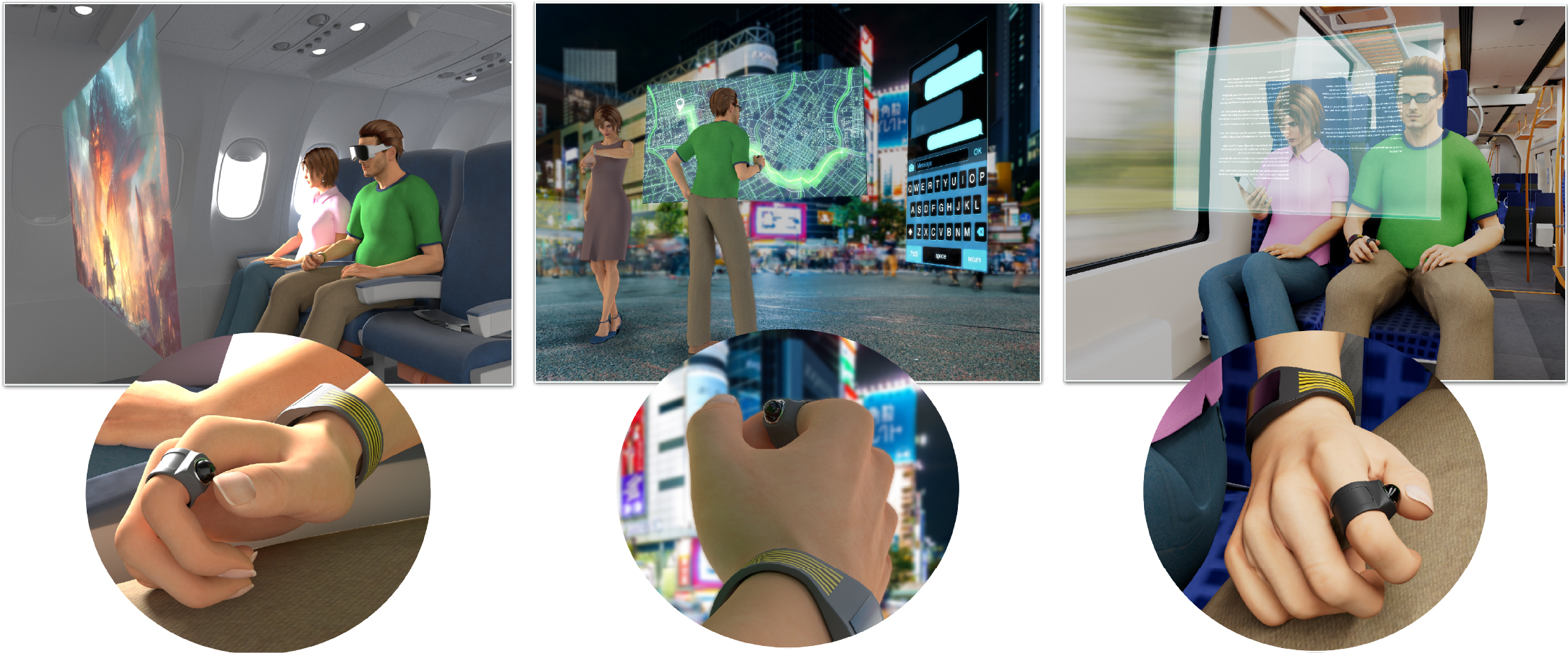}
  \caption{Application examples of picoRing \textit{mouse}.}
  \label{fig:app}
  \Description{Figure 8 illustrates application examples of the picoRing mouse. The images depict three scenarios: on an airplane, a person using the device to interact with a projected display; in a cityscape, a user interacting with a virtual map and keyboard; and on public transport, a person using the device to read a virtual document. Each scenario includes a close-up view of the picoRing mouse on the user's finger, highlighting its integration with wearable technology.}
\end{figure*}

\subsection{SNR of picoRing \textit{mouse}}
We then evaluate the signal-to-noise ratio (SNR) of picoRing \textit{mouse} for user's scrolling and pressing interactions, input power, distance between the ring and wristband, the finger bending, and public environments.
Similar to \cite{takahashi_picoring_2024}, the SNR was calculated as follows using S21 $100$ outputs~$\left(P_{\rm out}~(\si{\dB}) = 20\log_{\rm 10} V_{\rm out} \right)$ of the VNA:
\begin{align*}
    \mbox{SNR}(\mbox{a.u.}) = \cfrac{\mbox{mean}~\left(P_{\rm out~w/~ring}\right) - \mbox{mean}~\left(P_{\rm out~w/o~ring}\right)}{\mbox{std}~\left(P_{\rm out~w/o~ring}\right)} \label{eq:snr}
\end{align*}
Here, we consider that the SNR over $10$ can support high-fidelity recognition over $99\%$ accuracy, referring to \cite{takahashi_picoring_2024}. 
Also, we decouple the readout board with the common ground using the RF transformers~(T1-1-X65+, Mini Circuits).
\autoref{fig:snr_in_controlled}a shows the time-series SNR and the average SNR for the pressing and 2D scrolling inputs.
The average SNR was calculated using data collected over a 5-minute period from three invited participants with almost similar hand sizes (one woman and two men, 20s).
The result shows that the average SNR for each input is over \SI{10}{\dB} with minimal variations, indicating that picoRing \textit{mouse} can support high-fidelity recognition of thumb-to-mouse finger inputs for a similar hand size.
To be available for various hand sizes, we will quantitatively explore how the coil size can affect the SNR using those like electromagnetic simulators.

Next, we measured how the distance and orientation of the ring coil influence the SNR of picoRing \textit{mouse}.
Original picoRing~\cite{takahashi_picoring_2024} have shown that the SNR characteristics is almost the same with or without the hands, thanks to the inductive inherent robustness to dielectric human presence.
Therefore, we conducted this evaluation without human hands to easily adjust the distance and orientation between the coils.
Specifically, we prepared a set of jigs to change the distance and orientation of the ring coil by the 3D printer (see \autoref{fig:snr_in_controlled}b).
The range of he distance is from \SI{10}{\cm} to \SI{20}{\cm} in step of \SI{1}{\cm} in addition to varying the angle of the ring coil from \SI{-30}{\degree} to \SI{60}{\degree} in step of \SI{30}{\degree}.
We prepared the two types of ring coils: straight-type coil and tilted-type coil with the coil angled about \SI{20}{\degree} to the vertical axis of the ring.
Because the thumb-to-index input, which is typically performed with the finger in the bent position, can cause misalignment between the ring and wristband coils, weakening magnetic coupling, the tilted ring with the coil angled to the vertical axis enables to maintain the alignment and strengthen the inductive coupling.
Note that the ring's resonant frequency is tuned at \SI{28}{\MHz}.
The result shows that the available distance is estimated to be approximately \SI{14}{\cm} because the SNR is over $10$ against up to \SI{60}{\degree} finger bending.
These available distance and angle provided by picoRing \textit{mouse} are well-suited for the reliable thumb-to-index input during the typical middle-sized hand movements.
Furthermore, the tilted ring could increase the SNR by approximately $3$ when the finger is bent by around \SI{30}{\degree} because the measured inductive coupling coefficient~($k$) increases from $0.0031$ to $0.0039$.

Then, we evaluated the SNR varying the input power of the wristband coil from \SI{-30}{dBm} ($=\SI{1}{\uW}$) to \SI{10}{dBm} ($=\SI{10}{\mW}$) in steps of \SI{5}{\dB}~(see \autoref{fig:snr_in_controlled}c).
The distance and angle between the ring and the wristband is set to \SI{14}{\cm} and \SI{30}{\degree}, respectively.
The result indicates the input power ranging from \SI{-15}{dBm} (=\SI{30}{\uW}) to \SI{5}{dBm} (=\SI{3.2}{\mW}) achieves the SNR over $10$, although the high input power over \SI{10}{dBm} ($=\SI{10}{\mW}$) unfortunately causes the decrease of SNR due to the fluctuation of the amplifier in the bridge circuit.
To save the operational power of the bridge circuit, we chose \SI{-7}{dBm} ($=\SI{0.2}{\mW}$) as the input power of the NanoVNA, from the available power options in the NanoVNA.

Finally, we measured the SNR during the daily usage.
Although the SNRs for proximity to metallic items and electrical appliances are shown to be robust in picoRing~\cite{takahashi_picoring_2024}, our focus is on measuring the SNR against electromagnetically-noisy situations in real-world wearable computing scenarios.  
Here, we consider the following three situations: using the ring while 1) seated in a train, 2) standing in crowded areas in addition to being surrounded by the multiple smartphones, and 3) driving a car.
Note that the measured noise level in these situations is approximately \SI{20}{dBm\per\Hz} higher than in typical environments.
\autoref{fig:snr_in_real} shows the time-series SNR spectrum and average SNR for these situations, based on data from a single user~(20s, man).
The result indicates that picoRing \textit{mouse} successfully detects a sharp peak at \SI{28}{\MHz} with the sufficient SNR over $25$ despite proximity to the metallic items~(\textit{e.g.,} steering wheels) and electrical appliances (\textit{e.g.,} laptops, smartphones, electronic control units in the cars and trains).

\section{APPLICATION EXAMPLES}

picoRing \textit{mouse} offers a significant advantage in providing long-term continuous operation on a single charge, while reliably detecting subtle thumb-to-index inputs. 
Therefore, its primary application is ubiquitous finger-based interaction systems, particularly when integrated with augmented reality (AR) glasses or head-mounted displays (HMD)~(see \autoref{fig:app}). 
AR glasses and HMDs often rely on dynamic hand gestures for scrolling interactions, which can lead to fatigue over time. 
In contrast, picoRing \textit{mouse} offers subtle finger inputs that enable comfortable, long-term interactions while preserving user privacy. 
This makes it an ideal solution for seamless engagement with digital content.
Additionally, the device's efficient power management allows frequent and natural use without concerns about battery life, enhancing its practicality in everyday AR and HMD applications.
\autoref{fig:app} shows some application examples of picoRing \textit{mouse}.
The discreet nature of ring inputs are not easily noticed by the surroundings, minimizing any disruption to those nearby.
For example, in an airplane or train, a user discreetly interacts with a personal screen using the picoRing \textit{mouse}, allowing for private control without disturbing nearby passengers.
Even in an urban crowded space, users navigate a large public display or AR map with the picoRing \textit{mouse}, seamlessly interacting with digital content (\textcolor{blue}{\url{https://youtu.be/O7E_RQ8KSLQ}}).

\begin{figure*}[ht!]
  \centering
  \includegraphics[width=1.0\textwidth]{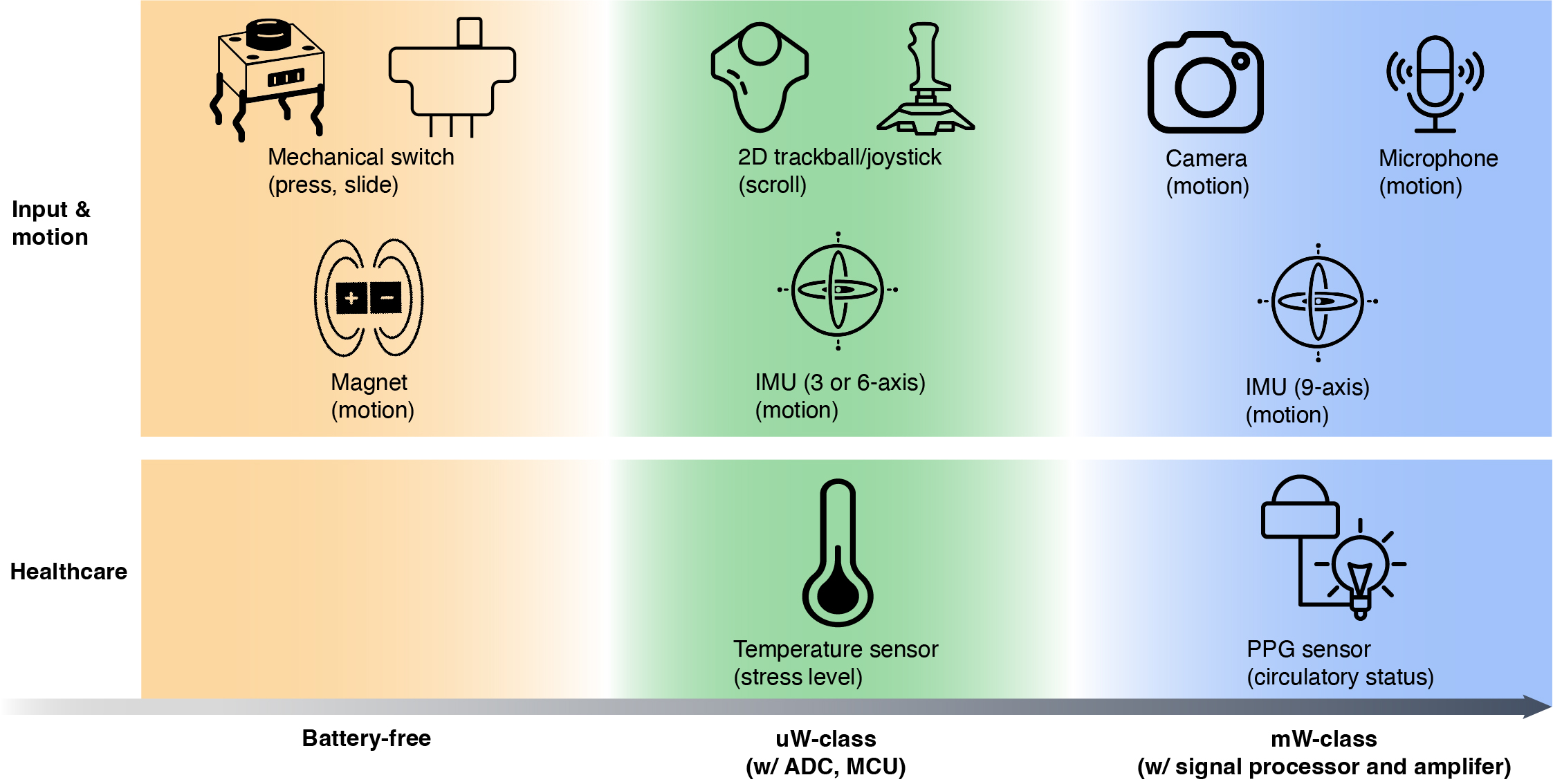}
  \caption{Examples of ring-sized small sensors for input \& motion or healthcare purpose.}
  \label{fig:design_space}
  \Description{This figure presents ring-sized sensors for input, motion, and healthcare, categorized by power consumption: battery-free, uW-class, and mW-class. For input and motion, battery-free sensors include a mechanical switch for press and slide actions and a magnet for motion detection. The uW-class features a 2D trackball/joystick for scrolling and a 3 or 6-axis IMU for motion sensing. The mW-class includes a camera and microphone for motion detection, along with a 9-axis IMU. For healthcare, the uW-class has a temperature sensor for stress level monitoring, while the mW-class includes a PPG sensor for circulatory status assessment. The layout highlights the power requirements and functionalities of each sensor type.}
\end{figure*}

\section{DISCUSSION}\label{sec:discussion}

\textit{Dual-device requirement.} 
picoRing \textit{mouse} requires users to attach the two types of wearable devices---ring and wristband---unlike the standard wearables like smartwatches and smart rings.
Given that many users already wear wristbands throughout the day for health and fitness tracking, as well as for convenient visual displays, the burden of wearing an additional ultra-low-power ring that requires minimal charging is quite light, making it easy to incorporate wearing a pair of the ring and the wristband into daily routines.

\textit{Bulky and power-consuming readout board.} 
Currently, our readout board, which includes the bridge, VNA, and the external PC, requires a smartphone-sized NanoVNA. 
However, the board could be miniaturized to fit into a smartwatch. 
NFC, which hardware is optimized for the coil-based communication~\cite{takahashi_full-body_2025}, could decrease the total power consumption less than \SI{50}{\mW}~\cite{kg_near_nodate,zhao_nfc-wisp_2015}.
Such a low power operation of the wristband coil could allow the integration of our wristband system into the commercially-available smartwatch, which is typically driven by about \SI{500}{mAh}-class Lipo battery~\cite{gonzalez-canete_feasibility_2021}.

\textit{Wireless charging while in use.}
Wearable devices often recharge naturally when not in use, such as earbuds placed in dedicated charging cases. 
However, rings and wristbands are typically worn continuously, making this more challenging. 
The integration of a wearable power transmission system such as charging-enabled textiles could achieve the continuous power supply to picoRing \textit{mouse} in the use~\cite{sato_friction_2025,takahashi_meander_2022,li_plug-n-play_2025}.
This would ensure that picoRing \textit{mouse} remains powered throughout daily activities, enabling to be worn continuously as a fashion accessory.

\textit{Limited user study.} 
In the current study, we primarily focused on measuring the SNR related to mouse input. 
Unlike antenna-based wearable devices operating at higher frequencies (e.g., 2.4 GHz or 920 MHz) which are significantly affected by the human body's high dielectric loss, our coil-based system experiences almost no electromagnetic interference from a person. 
This means the signal-to-noise ratio (SNR) and frequency response remain largely unchanged whether the device is worn or not by human hand.
One potential user study will be a multi-factorial repeated-measures ANOVA to assess additional aspects such as ease of use, social acceptance, and fatigue~\cite{chen_efring_2022}. 
These evaluations will provide further understanding of the picoRing \textit{mouse} in various real-world scenarios, helping to refine its design and improve user experience.

\textit{Extension of ring functionality.}
picoRing \textit{mouse} has the potential to become a versatile ultra-low-powered wearable platform for a wide range of applications by integrating various low-powered sensors.
For instance, by incorporating IMUs~\cite{shen_mousering_2024,zhao_single_2022} or speaker~\cite{yu_ring--pose_2024}, picoRing can be transformed into motion sensing rings (see \autoref{fig:design_space}).
Additionally, integrating PPG or temperature sensors could transform it into a healthcare ring.
Furthermore, wearing picoRing \textit{mouse} on both hands can enable wearable VR/AR interface even in public spaces, by converting the user’s subtle mouse input to extended 3D hand dynamic posture~\cite{kari_handycast_2023}.

\textit{Optimization for further ring's low-powered operation.}
We are utilizing the ultra-low-powered MCU (STM32L011F4U6, STM), and there is still room for reducing power consumption in both hardware and software aspects. 
For instance, we can lower the operating voltage from \SI{1.8}{\V} to \SI{1.2}{\V} or reduce the MCU's clock speed. 
We would explore these options to increase further battery runtime, resulting in a decrease of the battery size.

\textit{Error handling in frequency-shift keying.}
The current frequency-shift keying in our ring-to-wristband communication is so simple, using different frequencies to send the digital information from the mouse sensor.
This simplicity presents several challenges, such as security concerns and data integrity issues. 
Furthermore, the slow data rate of $10-200$ bps in the current picoRing \textit{mouse} allows the high SNR against electromagnetic noise from nearby devices. 
However, the fast communication at $10-200$ kbps-level such as for IMU data would make picoRing \textit{mouse} vulnerable to electromagnetic noise, causing a significant drop in SNR.
To ensure reliable and fast ring-to-wristband communication, error handling techniques such as forward error correction would be commonly available.

\section{CONCLUSION}

This paper presents picoRing \textit{mouse}, an ultra-low-power ring-based finger input device. 
By combining semi-PIT-based ring-to-wristband low-powered wireless communication with the low-powered mouse module, picoRing \textit{mouse} achieves long-term operation of approximately 600 (8hrs use/day)-1000 (4hrs use/day) hours, in addition to supporting reliable, multi-modal, subtle inputs.
Our technical evaluation shows the strong robustness of the ring-to-wristband wireless link against different users, hand postures, and potential situations in the field of wearable computing.
We strongly believe picoRing \textit{mouse} could provide a new class of long-term wearables, seamlessly interconnecting itself with daily HCI devices.

\begin{acks}
This work was supported by JST ACT-X JPMJAX21K9, JSPS KAKEN 22K21343, JST ASPIRE JPMJAP2401, and Asahi Glass Foundation. 
\end{acks}

\bibliographystyle{ACM-Reference-Format}
\bibliography{references}

\appendix
\end{document}